\documentclass[12pt]{article}
\usepackage{amsmath}
\usepackage{amsfonts}
\usepackage{float}
\usepackage{booktabs}
\usepackage{graphicx}
\usepackage{caption}
\usepackage{multirow}
\usepackage{subcaption}
\usepackage{natbib}
\usepackage{url}
\usepackage{array}
\usepackage{longtable}

\newcommand{\blind}{0}

\begin{document}

\def\spacingset#1{\renewcommand{\baselinestretch}%
{#1}\small\normalsize} \spacingset{1}

\newcommand{\mbf}[1]{\mathbf{#1}}
\newcommand{\mbv}[1]{\mbox{\boldmath$#1$\unboldmath}}

\def\ba{\mathbf{a}}
\def\bb{\mathbf{b}}
\def\bc{\mathbf{c}}
\def\bd{\mathbf{d}}
\def\be{\mathbf{e}}
\def\bh{\mathbf{h}}
\def\bht{\tilde{\mathbf{h}}}
\def\bi{\mathbf{i}}
\def\bj{\mathbf{j}}
\def\bk{\mathbf{k}}
\def\bo{\mathbf{o}}
\def\bp{\mathbf{p}}
\def\bq{\mathbf{q}}
\def\br{\mathbf{r}}
\def\bs{\mathbf{s}}
\def\bt{\mathbf{t}}
\def\bu{\mathbf{u}}
\def\bv{\mathbf{v}}
\def\bw{\mathbf{w}}
\def\bx{\mathbf{x}}
\def\by{\mathbf{y}}
\def\bz{\mathbf{z}}
\def\bA{\mathbf{A}}
\def\bB{\mathbf{B}}
\def\bC{\mathbf{C}}
\def\bD{\mathbf{D}}
\def\bG{\mathbf{G}}
\def\bH{\mathbf{H}}
\def\bI{\mathbf{I}}
\def\bK{\mathbf{K}}
\def\bM{\mathbf{M}}
\def\bP{\mathbf{P}}
\def\bQ{\mathbf{Q}}
\def\bR{\mathbf{R}}
\def\bS{\mathbf{S}}
\def\bT{\mathbf{T}}
\def\bV{\mathbf{V}}
\def\bU{\mathbf{U}}
\def\bW{\mathbf{W}}
\def\bX{\mathbf{X}}
\def\bY{\mathbf{Y}}
\def\bZ{\mathbf{Z}}
\def\bzero{\mathbf{0}}
\def\bfone{\mathbf{1}}
\newcommand{\bftheta}{\mbox{\boldmath $\theta$}}
\newcommand{\bfTheta}{\mbox{\boldmath $\Theta$}}
\newcommand{\bfgamma}{\mbox{\boldmath $\gamma$}}
\newcommand{\bfalpha}{\mbox{\boldmath $\alpha$}}
\newcommand{\bfeta}{\mbox{\boldmath $\eta$}}
\newcommand{\bfbeta}{\mbox{\boldmath $\beta$}}
\newcommand{\bfzeta}{\mbox{\boldmath $\zeta$}}
\newcommand{\bflambda}{\mbox{\boldmath $\lambda$}}
\newcommand{\bfdelta}{\mbox{\boldmath $\delta$}}
\newcommand{\bfmu}{\mbox{\boldmath $\mu$}}
\newcommand{\bfnu}{\mbox{\boldmath $\nu$}}
\newcommand{\bfomega}{\mbox{\boldmath $\omega$}}
\newcommand{\bfell}{\mbox{\boldmath $\ell$}}
\newcommand{\bfepsilon}{\mbox{\boldmath $\varepsilon$}}
\newcommand{\bfsigma}{\mbox{\boldmath $\sigma$}}
\newcommand{\bfPhi}{\mbox{\boldmath $\Phi$}}
\newcommand{\bfPsi}{\mbox{\boldmath $\Psi$}}
\newcommand{\bfphi}{\mbox{\boldmath $\phi$}}
\newcommand{\bfpsi}{\mbox{\boldmath $\psi$}}
\newcommand{\bfSigma}{\mbox{\boldmath $\Sigma$}}
\newcommand{\bfLambda}{\mbox{\boldmath $\Lambda$}}
\newcommand{\bfDelta}{\mbox{\boldmath $\Delta$}}
\newcommand{\bfxi}{\mbox{\boldmath $\xi$}}
\def\ave{\textrm{ave}}
\def\var{\textrm{var}}
\def\cov{\textrm{cov}}
\def\diag{\textrm{diag}}
\def\trace{\textrm{tr}}
\def\pr{\textrm{Pr}}
\def\VAR{\textrm{VAR}}
\def\Gau{\textit{Gau}}

%%%%%%%%%%%%%%%%%%%%%%%%%%%%%%%%%%%%%%%%%%%%%%%%%%%%%%%%%%%%%%%%%%%%%%%%%%%%%%

\if0\blind
{
  \title{\bf Locally Adaptive Shrinkage Priors for Trends and Breaks in Count Time Series}
  \author{Toryn L. J. Schafer\thanks{
    The authors gratefully acknowledge \textit{please remember to list all relevant funding sources in the unblinded version}}\hspace{.2cm}\\ 
    Department of Statistics, Texas A\&M University\\
    and \\
    David S. Matteson\\
    Department of Statistics and Data Science, Cornell University}
  \maketitle
} \fi

\if1\blind
{
  \bigskip
  \bigskip
  \bigskip
  \begin{center}
    {\LARGE\bf Locally Adaptive Shrinkage Priors for Trends and Breaks in Count Time Series}
\end{center}
  \medskip
} \fi

\bigskip
\begin{abstract} %currently 180 words
    Non-stationary count time series characterized by features such as abrupt changes and fluctuations about the trend arise in many scientific domains including biophysics, ecology, energy, epidemiology, and social science domains. Current approaches for integer-valued time series lack the flexibility to capture local transient features while more flexible models for continuous data types are inadequate for universal applications to integer-valued responses such as settings with small counts. We present a modeling framework, the negative binomial Bayesian trend filter (NB-BTF), that offers an adaptive model-based solution to capturing multiscale features with valid integer-valued inference for trend filtering. The framework is a hierarchical Bayesian model with a dynamic global-local shrinkage process. The flexibility of the global-local process allows for the necessary local regularization while the temporal dependence induces a locally smooth trend. In simulation, the NB-BTF outperforms a number of alternative trend filtering methods. Then, we demonstrate the method on weekly power outage frequency in Massachusetts townships. Power outage frequency is characterized by a nominal low level with occasional spikes. These illustrations show the estimation of a smooth, non-stationary trend with adequate uncertainty quantification.
\end{abstract}

\noindent%
\vfill

\newpage
\spacingset{2} % DON'T change the spacing!
\section{Introduction}
\label{sec:intro}

The methodological development for sequentially observed count valued data has been ongoing for decades without an established generic method to date \citep{davis2021count, davis2016handbook}. The success of established methods for continuous data such as the autoregressive moving average model motivated initial analyses of integer-valued time series as continuous processes (possibly after transformation). The approximate model is suitable for untransformed integer-valued time series characterized by large counts, but it is not generalizable to cases especially those characterized by small counts including zeros. Additionally, a model for continuous data cannot generate integer-valued responses invalidating inference for generated quantities (i.e., predictions and forecasts) and highlighting inherent model misspecification \citep{kowal2020simultaneous}. A typical strategy for handling zero counts is to add an offset then a log-transformation, but the offset term must be considered as an additional tuning parameter chosen with respect to the scale of the data as the value of the offset can lead to the zero integer observation being considered an outlying value in the transformed space. Ultimately, development of methods specific for integer-valued data is needed for more general use by the scientific community.

Non-stationary count time series characterized by features such as abrupt changes and fluctuations about the trend arise in many scientific datasets collected in domains such as biophysics \citep{losey2022simulating}, epidemiology \citep{palmer2021count, cazelles2018accounting}, emergency services \citep{matteson2011forecasting}, and energy, the later introduced application area. It is often of more interest to investigate the underlying trend in the presence of heterogeneity and multi-scale features in the observations. The characterization of drift in a time series describes micro-scale evolution of a process appearing as variation about a gradual trend such as seasonal fluctuations. At the same time, we want a framework to handle abrupt shifts causing macro-level changes in a process. The presence of drift and shift in applications is challenging for strict modeling frameworks that assume specific parametric forms or global smoothness of the mechanisms of the dynamic process. Our aim is to estimate a locally smooth trend in the presence of such macro level changes (abrupt spikes) and micro level fluctuations. 

A naive method traditionally used for filtering a temporal trend is exponential smoothing. Exponential smoothing estimates a trend by a weighted average of past values. Simple exponential smoothing is a weighted average filtering estimate where the weights decay exponentially such that observations nearby in time contribute more to the estimate than farther in the past \citep{hyndman2008forecasting}. The lack of parametric structure allows the method to estimate trends in a variety of time series characterized by drift and shift. However, the estimates tend to be characterized by a lot of wiggliness (not smooth) as it tends to chase noise in a time series. The lack of smoothness can be made even more apparent by the discrete nature of the count data. It is often of interest to estimate smoother trends than those provided by exponential smoothing yet retain the flexibility necessary in non-stationary cases. Our modeling framework offers an adaptive model-based solution to the analysis of sequentially observed count data with multiscale features. The underlying time-varying signal is estimated by a generalization of the computationally efficient dynamic shrinkage prior for continuous time series \citep{kowal2019dynamic}. 

Our model falls into the current popular approach for non-stationary count time series, state space modeling. In state space modeling, non-stationarity results from the dynamics of random effects in the latent space and trend dynamics are separated from observational heterogeneity \citep{durbin2000time, aktekin2018sequential, fruhwirth2006auxiliary}. Numerous state-space model specifications exist, but most traditional choices lack locally adaptive smoothness. We take a novel approach to the specification of the latent process by including dependent shrinkage through a global-local shrinkage prior on the volatility of a differenced time series rather than specifying the dynamic process on the latent conditional mean. For computational efficiency, we chose the negative binomial data model that is sampled by a conjugate update resulting from the P{\'o}lya-Gamma (PG) parameter expansion technique of \citet{polson2013bayesian}. Computational efficiency and simple default prior choices ensures the ease of use by practitioners of the presented framework. 

Our framework is the first to present the dynamic shrinkage process for count time series. There exist few works that contain a subset of the aspects we present. A shrinkage prior for sparse counts was developed by \citet{datta2016bayesian} and shrinks the absolute value of the trend without temporal dependence. The aforementioned shrinkage prior as presented cannot be easily generalized to shrinkage of the increments (differenced trend) as it is a strictly positive prior distribution. \citet{palmer2021count} analyze the non-stationarity of COVID-19 case counts by second differencing the latent trend and defining a parametric evolution for the differences. Alternatively, we define a continuous shrinkage prior for the differenced trend and allow for evolution to occur in the shrinkage process, i.e., the amount of regularization of the increments are correlated rather than the actual expected value of the increments. \citet{tibshirani2014adaptive} describes a penalized non-parametric regression formulation of trend filtering that is an extension to the one-dimensional fused lasso \citep{tibshirani2005sparsity}.  The lasso penalty regularizes the differences in the coefficients analogous to our model but without temporal dependence in the shrinkage.

A motivating application for our framework is the frequency of power outages in the electric energy grid. Power outages are an indicator of the resiliency of the grid system and understanding mechanisms that effect resiliency can help planners foresee potential risks \citep{checastaldo2021critical}. We use publicly available data from the state of Massachusetts on power outage events. The data has been previously presented in an analysis of grid resiliency to animal related outages \citep{feng2022analysis}. Our proposed framework may be used for future analyses of associations between power outage trends and hypothesized causes of disruptions to the power grid.

The remainder of the paper is outlined as follows. First, Section \ref{sec_prelim} reviews a foundational model in count time series, Poisson state-space model. Section \ref{sec_model} describes the components of our proposed Bayesian hierarchical model: the observation model \ref{sec_obs}, the trend evolution model \ref{sec_evol}, and the remaining prior specifications \ref{sec_param}. Next, Section \ref{sec_samp} details the MCMC sampling steps that are a combination of Gibbs, accept-reject, and slice sampling with complete derivations of conditional updates between the text and Supplement. Then a simulation study evaluates the method performance in Section \ref{sec_sim} against several alternatives and an illustration of the method on the Massachusetts power outage data is presented in Section \ref{sec_power}. Finally, we wrap up with some discussion and future work in Section \ref{sec_disc}. Additional technical details of the MCMC sampling algorithm and performance summaries of models in simulation are provided in the Supplement.

\section{State Space Models for Count Time Series}\label{sec_prelim}

Count time series models are generally classified into two types: parametric or observation driven \citep{cox1981statistical}. The classification is determined by whether the dynamics evolve in a specified latent process (parametric driven) or as a function of past observed values (observation driven). Our model falls into the parametric driven class as the evolution occurs in a latent process and the observations are subsequently conditionally independent given the process. Specifically, we use a model in the class of generalized state-space models. 

We use a relaxed definition of the state-space model with two basic components: the data generating model given a latent process and the evolution model of the latent process. Moreover, the conditioning on the latent process leads to conditional independence among the observations. We focus our review on a Poisson state-space model with a random walk process on the first differences. Let the univariate, integer-valued observations be denoted $Y_{t}$ for $t = 1, ..., T$ then the observation and state equations are:
\begin{equation}\label{eq:ssm}
    \begin{gathered}
        Y_t \vert \theta_t \sim \textrm{Pois}(e^{\theta_t}), \\
        \theta_t = \theta_{t-1} + \omega_t, \quad \omega_t \overset{\mathrm{IID}}{\sim} \text{N}(0, \sigma^2)\\
    \end{gathered}
\end{equation}
where $\theta_t \in \mathbb{R}$ is the state at time $t$ that evolves according to a random walk, with i.i.d. mean zero first order increments, $\omega_t$, from a Gaussian distribution, for example. A model for the initial state values is specified to begin the recursion. 

\begin{figure}
    \centering
    \includegraphics[width=0.95\textwidth]{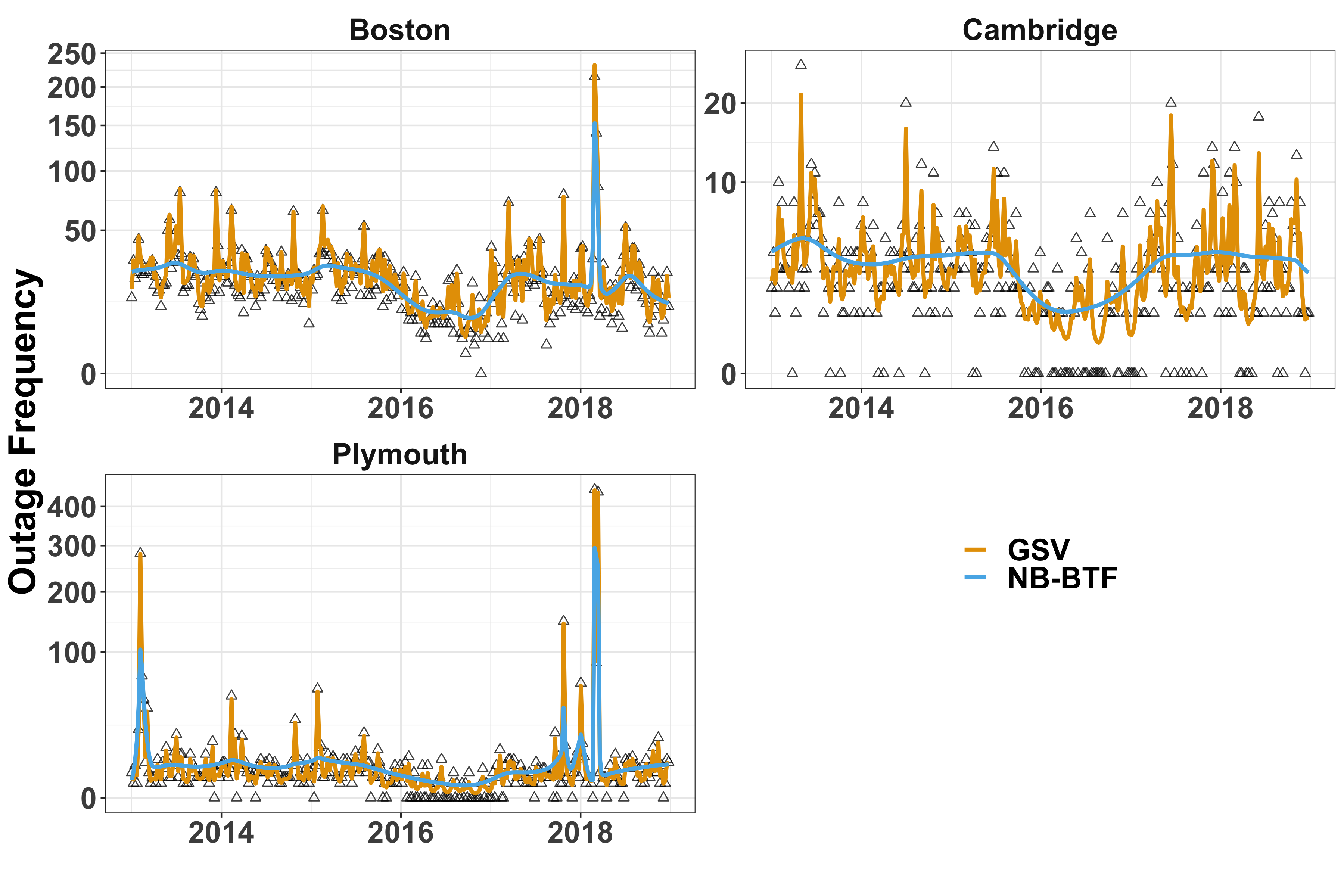}
    \caption{A comparison of the trend estimates for weekly power outage frequency in three Massachusetts townships using a Poisson observation model with Gaussian stochastic volatility for the first increments (GSV; orange line) and the proposed negative binomial with dynamic horseshoe Bayesian trend filter observation model (NB-BTF; blue line). It is clear the GSV model overfits to the observed data (x's) in all three scenarios while the regularization on the increments of NB-BTF produces a locally smooth trend estimate. The y-axis is square-root transformed.}
    \label{fig:bench}
\end{figure}

A first extension of \eqref{eq:ssm} is to include stochastic volatility in the evolution equation \citep{kim1998stochastic}. This would be applicable to cases where the variance of the increments is heteroscedastic, i.e., the rate of change in the mean exhibits non-constant variance. The evolution process with Gaussian stochastic volatility on the first order increments, $\omega_t$, can be expressed as the following product process:
\begin{equation}\label{eq:sv1}
    \begin{gathered}
        \omega_t = \exp(h_t/2) e_t, \quad e_t \overset{\mathrm{IID}}{\sim} \text{N}(0,1),\\
        h_t = \mu + \phi(h_{t-1} - \mu) + \eta_t, \quad \eta_t \overset{\mathrm{IID}}{\sim} \text{N}(0, s^2),\\
    \end{gathered}
\end{equation}
where $e_t$ is standard Gaussian noise, $h_t$ evolves according to a first order autoregressive evolution with mean $\mu$, autocorrelation $\phi$, and innovations $\eta_t$ from a mean zero Gaussian distribution with standard deviation $s$ independent of $e_t$. It is worth noting that given the definition of the increments in \eqref{eq:sv1}, $\exp(h_t/2)$ is the (conditional) standard deviation of $\theta_t$ in \eqref{eq:ssm}. Stochastic volatility describes a broad set of models, but primarily they are characterized by a lack of locally adaptive smoothness. While models \eqref{eq:ssm} and \eqref{eq:sv1} assume mean zero for the increments, $\omega_t$, there is no additional noise regularization leading to potential overfitting of noisy observations as seen in Figure \ref{fig:bench}. Specifically, in the next section, we introduce our proposed data-adaptive solution to estimating smooth trends in the presence of data heterogeneity.
% below we're just changing the priors, so how does the prior shape play into the behavior...
% A latent Gaussian process (LGP) model with a Poisson data model is a specification \citep{holan2016hierarchical}. However, standard Gaussian processes typically are not adequate to capture non-stationarity in smoothness. 

\section{Negative Binomial Bayesian Trend Filter} \label{sec_model}

\begin{figure}[!ht]
    \centering
    \includegraphics{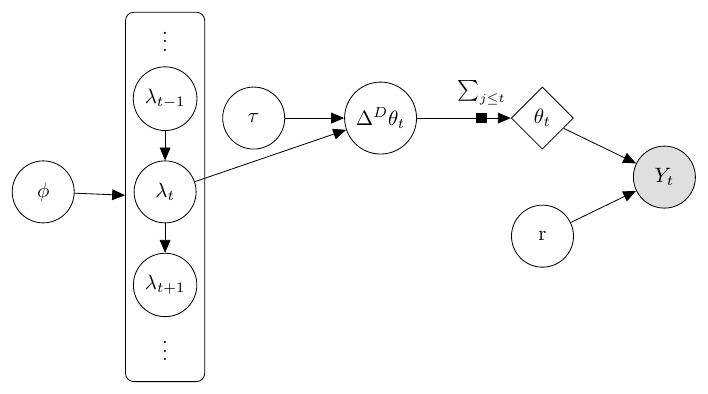}
    \caption{A directed graph cartoon of our proposed model structure focused on the time point $t$. The gray circle represents observed data while white circles represent unobserved latent values and parameters. The state $\theta_t$ encodes the latent log conditional mean of the data $Y_t$ and is derived by summation of the past latent increments and initial states as indicated by the summation and diamond shape. The temporal evolution of the process occurs among the local shrinkage parameters, $\lambda_t$, enclosed in the rectangle.}
    \label{fig:nb_dag}
\end{figure}

The following sections detail the components of the proposed negative binomial Bayesian trend filter (NB-BTF) for sequentially observed count data (Figure \ref{fig:nb_dag}). The model extends the framework presented in the previous section to account for overdispersion and local smoothing.

\subsection{The Observation Model} \label{sec_obs}

The Poisson distribution assumed in \eqref{eq:ssm} is limited by the relationship between the mean and variance. Often the data exhibit more heterogeneity than can be accounted for with the introduced stochastic volatility model \eqref{eq:sv1}. The stochastic volatility model captures heteroscedasticity in the rate of change in the mean, but does not allow for changes in observational variance beyond changes in the mean. In order to decouple the strict relationship between the mean and variance, we use a negative binomial distribution for the observations. Compared to the Poisson distribution, the additional parameter of the negative binomial distribution allows for modeling overdispersion in the variance of the observations. Formally, let the observations be denoted $Y_{t} \in \mathbb{N} = \{0, 1, 2, ..., \infty\}$ for $t = 1, ..., T$ then
\begin{equation} \label{eq:data}
    \begin{gathered}
        Y_{t} \vert \theta_{t}, r \sim \textrm{NB} \! \left( r, \frac{e^{\theta_{t}}}{r + e^{\theta_{t}}} \right), \quad \textrm{E}[Y_t \vert \theta_{t}, r] = e^{\theta_{t}},\\  \var{}(Y_t \vert \theta_{t}, r) = e^{\theta_{t}}(1 + e^{\theta_{t}}/r),
    \end{gathered}
\end{equation}
where $\theta_t$ for $t = 1, ..., T$ is the log of the conditional mean and $r > 0$ is a scalar overdispersion parameter. The conditional mean is defined the same as the Poisson state-space model in \eqref{eq:ssm}. Furthermore, the Poisson distribution is a limiting case as the overdispersion parameter increases, $r \rightarrow \infty$. In practice, the overdispersion parameter $r$ can be fixed at a numerically large constant to approximate a Poisson distribution without changing the sampling algorithm.

The likelihood of $\bY = (Y_1, Y_2, ..., Y_T)$ is
\begin{equation}\label{eq:likelihood}
L[\bY \vert \cdot] \propto  \prod_{t=1}^T \frac{(e^{\theta_{t} - \textrm{log} \, r})^{Y_{t}}}{(1 + e^{\theta_{t}- \textrm{log} \, r})^{Y_{t}+r}}. 
\end{equation}
The likelihood expression is used to establish the Bayesian sampling scheme with PG auxiliary variables as in \citet{polson2013bayesian}.

\subsection{The Evolution Process}\label{sec_evol}

% The model so far only allows for local smoothing (our new observation model and sv1). Global smoothing fails to capture multi-scale behavior observed in non-stationary time series. Local adaptivitity is necessary for data exhibiting regime changes in smoothness. It most useful in scenarios when ... 

As in \eqref{eq:ssm} and \eqref{eq:sv1}, we model the increments of the latent trend, $\omega_t = \Delta^D \theta_{t}$. Specifically, we use the computationally efficient dynamic horseshoe prior \citep[DHS;][]{kowal2019dynamic}.  The DHS is a flexible model for inducing temporally (i.e., locally) smooth trends.

The DHS falls into the class of global-local shrinkage priors with a global parameter to control the proportion of shrunk parameters across the time series and local parameters to allow locally in time unshrunk differences. Formally, there is a continuous scale mixture distribution on the volatility of the latent increments, $\bfomega$, parameterized by a global scale, $\tau$, and local scale parameters, $\lambda_t$, for $t = D+1,D+2,...,T$. Specifically,
\begin{equation} \label{eq:dsp}
    \Delta^D \theta_{t} \equiv \omega_t, \quad \omega_t \vert \tau, \lambda_t \sim \textrm{N}(0, \tau^2 \lambda_t^2).
\end{equation}
When comparing \eqref{eq:sv1} and \eqref{eq:dsp}, the conditional prior of $\omega_t$ remains a Gaussian scale mixture model. In the general case, the choice of prior for $\tau$ and $\lambda_t$ determines the amount of shrinkage and smoothness of the trend. 

The DHS induces a locally smooth trend by defining temporal dependence on the log volatilities of $\bfomega$:
\begin{equation} \label{eq:h_ar}
    h_t \equiv \log(\tau^2 \lambda_t^2) = \mu + \phi (h_{t-1} - \mu) + \eta_t, \quad \eta_t \overset{\mathrm{IID}}{\sim} Z(\alpha,\beta,0,1),
\end{equation}
where $\log(\tau^2) \equiv \mu$, $\log(\lambda_t^2) \equiv \phi (h_{t-1} - \mu) + \eta_t$, and $Z(\alpha,\beta,\mu_z,\sigma_z)$ is the Z distribution with density function
\begin{equation*} \label{eq:zdist}
    [z] = \{\sigma_zB(\alpha,\beta)\}^{-1}\exp\{(z-\mu_z)/\sigma_z\}^\alpha[1+\exp\{(z-\mu_z)/\sigma_z\}]^{-(\alpha+\beta)}, \, z \in \mathbb{R},
\end{equation*}
where $B(\cdot,\cdot)$ is the beta function. The evolution process, for $phi > 0$, induces persistence in the amount of regularization of nearby increments. Comparing to the stochastic volatility model \eqref{eq:sv1}, the global variance is now estimated from the data in \eqref{eq:dsp}. Additionally, the innovation distribution choice of \eqref{eq:h_ar} induces more smoothness as the increments are shrunk toward zero (Figure \ref{fig:bench}).

We can formally assess the shrinkage mechanism by the shape of the prior on the latent shrinkage proportion, $\kappa_t = 1/(1+\var(\omega_t \vert \tau, \lambda_t))$. The shrinkage parameter is interpreted as the proportion of shrinkage estimated for $\omega_t$; a value of 1 for the shrinkage proportion indicates the parameter will be shrunk to 0 and a value of 0 corresponds to unshrunk estimates. The prior on $\{\kappa_t\}$ induced by \eqref{eq:h_ar} when $\alpha = \beta = 1/2$ and $\phi = 0$ has the classic horseshoe shape as described by \citet{carvalho2010horseshoe}. As the autocorrelation, $\phi$, increases, the prior on the shrinkage increases in concentration around 0 and 1 suggesting the increments will be more smooth than a model using an independent horseshoe prior. Throughout, we take $\alpha = \beta = 1/2$ to place equivalent prior mass near 0 and 1 on $\kappa_t$. 

% Point out further generalizations and discussion are in Kowal?

The degree of differencing in trend filtering applications is a modeling choice typically determined by the amount of data available with higher degrees requiring more data. We explore models fit using second differences ($D=2$) as we apply the framework to situations with adequate data.  A non-zero value of the second differences,  $\omega_t = \Delta^2 \theta_{t} = (\theta_{t} - \theta_{t-1}) - (\theta_{t-1} - \theta_{t-2})$, implies a change in the trend (mean is changing) or a change in the increments of the trend (mean is changing at different rates). In our applications, we expect the latent trend to be subject to abrupt, transient changes implying the differenced trend, $\bfomega$, will be constant, near-zero with a small number of deviations around the time point of change. 

% The case $D = 0$, $\omega_t = \Delta^0 \theta_{t} = \theta_{t} $, may be of interest in specific applications, but we believe, in general, a constant mean trend is of less interest. An extension of the model to include an intercept term and/or covariates could relax the conditional constant mean assumption.

\subsection{Practical Implementation Details}\label{sec_param}
 The choice of prior for the global variance, $\tau$, is $\tau = \textrm{exp}\{\mu/2\} \sim \textrm{C}^{+}(0, \sigma_\tau)$. The scale value, $\sigma_\tau$, is related to the prior proportion of parameters expected to be shrunk \citep{piironen2017hyperprior}. We consider the default choice, $\sigma_\tau = 1$, of \citet{carvalho2010horseshoe}, but also consider the recommendation $\sigma_\tau = 1/\sqrt{T}$ of \citet{kowal2019dynamic} for models of heteroskedasticity in the observations; model \eqref{eq:data} exhibits non-constant variance due to the variance being a function of the non-constant mean, $\bftheta$. 
 
In order to ensure stationarity of \eqref{eq:h_ar}, the prior for the autocorrelation parameter, $\phi$, is $(\phi+1)/2 \sim \textrm{Beta}(10, 2)$ which constrains $\phi$ between $-1$ and $1$ and places majority of prior mass in the positive part of the support (prior mean of $2/3$ and prior mode of $4/5$).

For computational efficiency in the sampling under PG parameter expansion, we consider $r$ to be an integer valued parameter with prior $r \sim \textrm{Pois}(10)$ \citep{polson2013bayesian}. 

% Multiple priors were explored for the overdispersion parameter: $[r] \sim C^+(0,10)$, $[r] \sim \textrm{N}^+(0,10^2)$, $[r] \sim \textrm{Unif}(0,100)$, and $[r] \sim \textrm{Pois}(5)$. 

\section{Estimation and Inference}\label{sec_samp}

Estimation of the model parameters is non-trivial due to the non-linear relationships and dimension of latent variables. A Bayesian estimation scheme has some immediate advantages over an optimization scheme. Primarily, Bayesian sampling provides simultaneous uncertainty quantification for parameters and derived quantities such as the shrinkage proportion, $\kappa_t$. The uncertainty quantification in Bayesian sampling can be carried forward in subsequent models when trend filtering is a component of a larger analysis. We discuss further the issue of uncertainty quantification for the fused lasso in Section \ref{sec_sim}. 

Our model implementation uses MCMC as it required little tuning and is the standard in similar models. Sampling from the posterior of the parameters is a combination of Gibbs, Metropolis-Hastings, and slice sampling steps. The Gibbs sampling steps use the PG parameter expansion of \citet{polson2013bayesian}. In general, the posterior exhibits high amounts of autocorrelation due to the number of Gibbs sampling steps. The recommendation is to allow the sampler to run for many iterations of burn in on the order of 100,000 and additionally to thin the post-burnin samples.

The sampling algorithm pseudo-code for $k = 1,...,K$ MCMC samples is as follows:

\begin{itemize}
    \item \textit{Overdispersion:} sample $r$ defined in \eqref{eq:data} by integer random-walk Metropolis-Hastings with proposal \eqref{eq:prop_r}, 
    \item \textit{Trend:} sample $\bftheta$ defined in \eqref{eq:data} by Gibbs step with PG parameter expansion along with associated auxiliary parameters, $\bfxi^{\theta}$,
    \item \textit{Volatility:} sample $\bh$ defined in \eqref{eq:h_ar} by Gibbs step with approximation by mixture of Gaussians and PG expansion along with associated auxiliary parameters, $\bfxi^{\eta}$,
    \item \textit{Autocorrelation:} sample $\phi$ defined in \eqref{eq:h_ar} with slice sampling,
    \item \textit{Average volatility:} sample $\mu$ defined in \eqref{eq:h_ar} by Gibbs step with PG parameter expansion along with associated auxiliary parameter, $\xi^{\mu}$.
\end{itemize}

The details of sampling the DHS parameters (i.e., $\bh$, $\phi$, and $\mu$) have been discussed in \citet{kowal2019dynamic} and subsequent work \citet{wu2020adaptive} with original discussion of the volatility and autocorrelation sampling schemes in \citet{kastner2014ancillarity}. Our use of the negative binomial likelihood requires detailing the first two sampling steps associated with the overdispersion and trend parameters. We discuss the details of these steps below.

\subsection{Overdispersion}

There are a variety of choices for the sampling scheme of the overdispersion parameter, $r$, given the parameterization of the negative binomial we used \eqref{eq:data}. \citet{zhou2015negative} established a conjugate prior for an alternative negative binomial parameterization and therefore is not applicable here. Other alternatives to our sampling scheme is the slice sampler \citep{neal2003slice} and Metropolis-Hastings algorithm without the constraint to integer values. The constraint to integer values was chosen for improved computation time of the auxiliary PG variables for the trend, $\bftheta$ \citep{polson2013bayesian}. 

Specifically, the overdispersion parameter, $r$, is sampled using an integer random-walk Metropolis-Hastings step. The proposal distribution is a Discrete Uniform distribution over an integer grid centered on the current overdispersion value, $r^{(k)}$, for MCMC iteration $k$. The grid length is determined by a tuning parameter, $s$, that has a default value of 1. The minimum value of the grid is constrained to be greater than or equal to 1 in order to ensure finite, positive variance of the negative binomial distribution. Formally, the proposal distribution can be expressed as:
\begin{equation}\label{eq:prop_r}
    r^* \sim \textrm{DUnif}(\max\{r^{(k)} - s, 1\}, r^{(k)} + s),
\end{equation}
where $r^*$ is the proposed value of the overdispersion parameter for MCMC iteration $k+1$.

\subsection{State vector}

The sampling of the trend for a negative binomial likelihood \eqref{eq:data} has the advantage of being tuning free compared to sampling the model with a Poisson likelihood \eqref{eq:ssm}. Specifically, the state vector can be sampled in a block Gibbs step by using a PG parameter expansion of the likelihood and the ``all without a loop" algorithm of \citet{kastner2014ancillarity} as implemented by \citet{kowal2019dynamic} for sampling the continuous trend. The likelihood for the $t^{th}$ element of $\bftheta$ derived from \eqref{eq:likelihood} is
\begin{equation}
    \begin{split}\label{eq:theta_lik}
        L[\theta_{t} \vert \cdot] &\propto  \frac{(e^{\theta_{t} - \textrm{log}r})^{Y_{t}}}{(1 + e^{\theta_{t}- \textrm{log}r})^{Y_{t}+r}},\\
&\propto \int_0^\infty \textrm{exp}(-\xi^{\theta}_{t}/2(\theta_{t} - \textrm{log}r - 0.5(Y_{t} - r)/\xi^{\theta}_{t})^2)\text{p}(\xi^{\theta}_{t} \vert Y_{t}+r, 0)d\xi^{\theta}_t,
    \end{split}
\end{equation}
where the second line is a result of Theorem 1 of \citet{polson2013bayesian} and $\text{p}(\xi^{\theta}_{t} \vert Y_{t}+r, 0)$ is the density for a PG random variable with distribution $\xi^{\theta}_t \sim \textrm{PG}(Y_{t}+r,0)$. Furthermore, a Gaussian prior for $\bftheta$ is conditionally conjugate for the likelihood in \eqref{eq:theta_lik}. For a given degree of differencing, $D$, the prior \eqref{eq:dsp} can be expressed in matrix notation as $\textbf{D}^{(D)}\boldsymbol{\theta} = (\theta_1, ..., \theta_D, \omega_1, ..., \omega_{T-D})' \sim \textrm{N}(\boldsymbol{0}, \Sigma_\omega)$ where $\Sigma_\omega = \textrm{diag}(\{\sigma^2_t\}_{t=1}^T)$, and $\textbf{D}^{(D)}$ is the differencing matrix (i.e. a matrix of linear combinations). For $t = 1,...,D$, $\sigma^2_t$ is fixed at 100 while for $t > D$, $\sigma^2_t = \tau^2\lambda^2_t$. Therefore, the conditional update for $\bftheta$ is $\textrm{N}((\bQ_\theta)^{-1}(\Sigma_\xi^{-1})(\textrm{log}r\boldsymbol{1} + 0.5(\textbf{Y} - r\boldsymbol{1})'\Sigma_\xi), \bQ_\theta)$ where $\boldsymbol{1}$ is a $T$-dimensional column vector of ones,  \mbox{$\bQ_\theta = (\Sigma_\xi^{-1} + (\textbf{D}^{(D)})'\Sigma_\omega^{-1}\textbf{D}^{(D)})$}, and $\Sigma_\xi = \textrm{diag}(\{(\xi_{t}^{\theta})^{-1}\}_{t=1}^T)$.

The auxiliary variables, $\boldsymbol{\xi}^{\theta} = (\xi^{\theta}_1, ..., \xi^{\theta}_T)'$, are updated with the conditional PG distribution, $\boldsymbol{\xi}^{\theta} \sim \textrm{PG}(\textbf{Y} + r\boldsymbol{1}, \boldsymbol{\theta} - \textrm{log}r\boldsymbol{1})$. When $r$ is an integer, the elements of $\textbf{Y} + r\boldsymbol{1}$ are integers and sampling is more computationally efficient than non-integer values.

\section{Simulation}\label{sec_sim}

We explored the performance of the NB-BTF for simulated count time series. We created artificial count time series by constructing the continuous mean, $\exp\{\bftheta\} = \bfbeta$, as a modification of a doppler trend function and simulating from the negative binomial \eqref{eq:data} with known overdispersion parameter and mean. Therefore, our simulated observations are not generated from the proposed model, but generally are simulated to have the features of interest: small integer values and varying rates of smoothness over time. The impact of time series length is evaluated by varying the length of time series, $T \in \{200, 500\}$, through infilling data along the trend function. Additionally, the impact of variability in the counts is explored by varying the overdispersion parameter, $r \in \{1, 10, 1000\}$, where larger $r$ implies less variance for a given value of the mean and $r = 1000$ is approximately Poisson. We simulated 100 time series under each setting. 

We then examined the recovery of the known trend using the described NB-BTF, the continuous dynamic shrinkage model of \citet{kowal2019dynamic} on the counts (Gau-DHS), the continuous dynamic shrinkage on the transformed counts, $\log(Y_t + 1)$, (logGau-DHS), the fused lasso for Poisson regression \citep[Pois-FL;][]{tibshirani2014adaptive}, and exponential smoothing (Exp-Smooth). For the negative binomial model, $\sigma_\tau = 1$ and the variance of the initial components of $\bftheta$ were not sampled. Additionally under $r=1000$, we fixed the overdispersion parameter to $r = 1000$ to illustrate the utility of our method as an approximate Poisson model. For the Gaussian models, the default is used, $\sigma_\tau = \sigma_\epsilon/\sqrt{T}$, where $\sigma_\epsilon$ is the observation error in the linear model. All differencing methods were fit to the second order differences. For all Bayesian models, the MCMC algorithms were run for 105000 steps: the first 100000 were discarded as burnin and the remaining 5000 were thinned to a posterior sample of 1000. The Pois-FL model has one tuning parameter, the degree of penalty of the L1 penalty on the differenced coefficients. The parameter was chosen by cross-validation using 5 folds and the systematic sampling of the folds as used in the R package \textit{genlasso} and all Pois-FL models were fit using \textit{glmgen}. The exponential smoothing trend estimate was calculated using the default arguments in the R package \textit{forecast}. Additionally, uncertainty quantification from Pois-FL and Exp-Smooth requires some bootstrap estimation by resampling and therefore is not reported here due to the temporal dependence in the data.

We measured performance based on recovery of the true mean trend \eqref{eq:data} as it is the only common component across all models. For the transformation used in logGau-DHS, the regression coefficients correspond to an estimate of $\textrm{E}[\log(Y_t + 1) \vert \cdot]$. To obtain comparable estimates of the untransformed conditional expectation, we use the estimating equation $\textrm{E}[Y_t \vert \cdot] = \exp\{\beta_t + \sigma_\epsilon^2/2\} -1$ where $\beta_t$ is the linear regression coefficient and $\sigma_\epsilon$ is the observation error \citep{morrissey2020revisiting}. 

We calculated root-mean squared error (RMSE), mean credible interval width (MCIW), and empirical coverage for 95\% equal tail credible intervals for the trend, $\textrm{E}[Y_t \vert \cdot ] = \beta_t$:
\begin{equation*}\label{eq:rmse}
\begin{split}
        \textrm{RMSE}(\hat{\bfbeta}) = \sqrt{\frac 1T \sum_{t=1}^T (\beta_t - \hat{\beta}_t)^2}\\
        \textrm{MCIW}(\hat{\bfbeta}) = \frac 1T \sum_{t=1}^T (\hat{\beta}^{(97.5)}_t - \hat{\beta}^{(2.5)}_t)\\
        \textrm{Emp-Cov}(\hat{\bfbeta}) = \frac 1T \sum_{t=1}^T \textrm{I}(\hat{\beta}^{(97.5)}_t \geq \beta_t \geq \hat{\beta}^{(2.5)}_t)
\end{split}
\end{equation*} 
where for Bayesian methods, $\hat{\beta}_t$ is taken to be the median, $\hat{\beta}^{(97.5)}_t$ is the 97.5\% quantile, and $\hat{\beta}^{(2.5)}_t$ is the 2.5\% quantile of the posterior sample; $\textrm{I}(\cdot)$ denotes the indicator function. We report computation time, empirical coverage at additional nominal credible interval levels, and numeric summaries of each metric across all simulation settings in the Supplement.

\begin{figure}[ht!]
    \centering
    \includegraphics[width=0.95\textwidth]{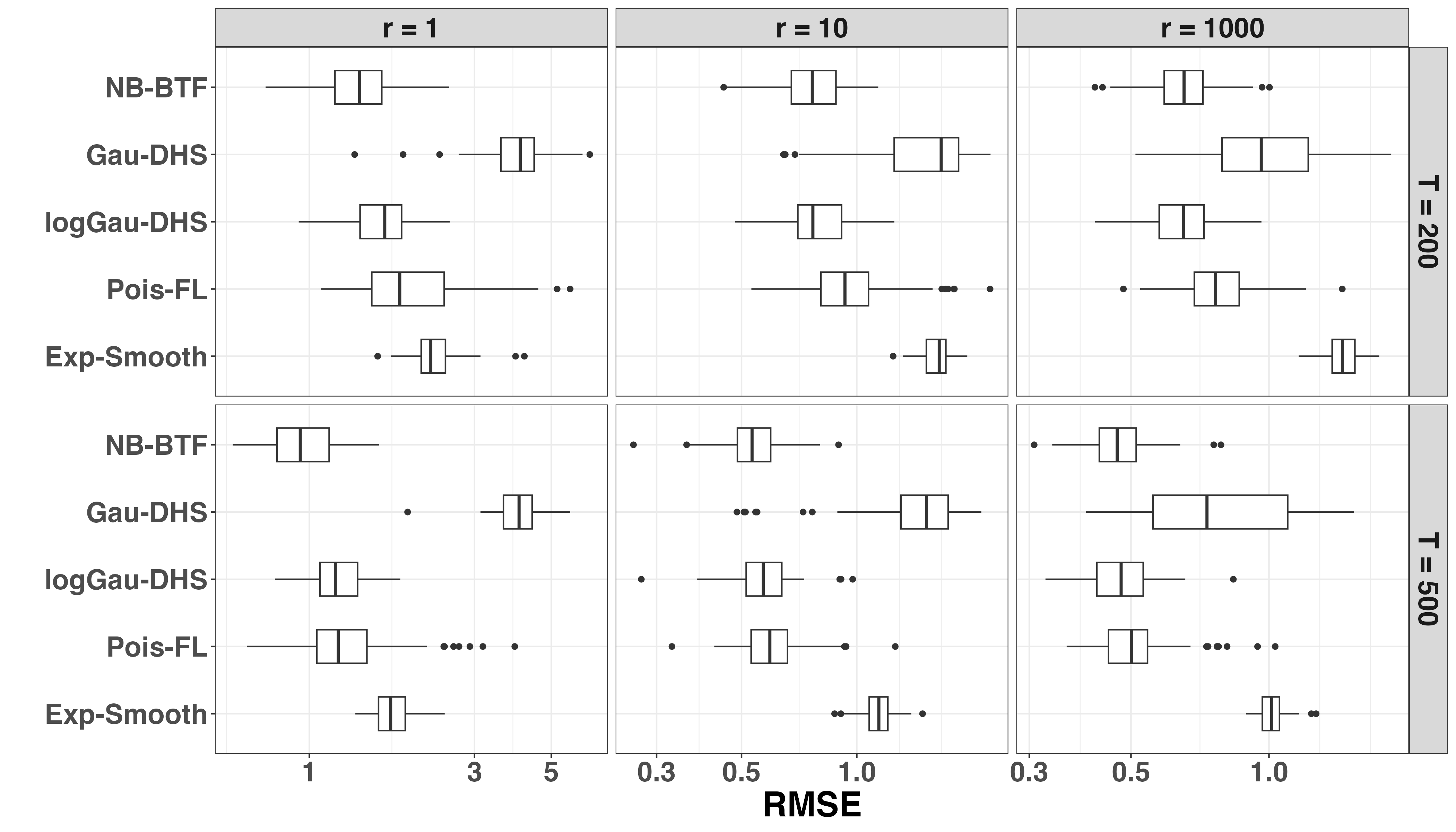}
    \caption{Boxplots of root mean squared error (RMSE) for estimation of the true trend under 100 simulations with varying time series length, $T$, and overdispersion, $r$. Five models were fit: NB-BTF, Gau-DHS, logGau-DHS, Pois-FL, and Exp-Smooth. The x-axis is transformed by log10.}
    \label{fig:sim_rmse}
\end{figure}

\begin{figure}[ht!]
    \centering
    \includegraphics[width = 0.95\textwidth]{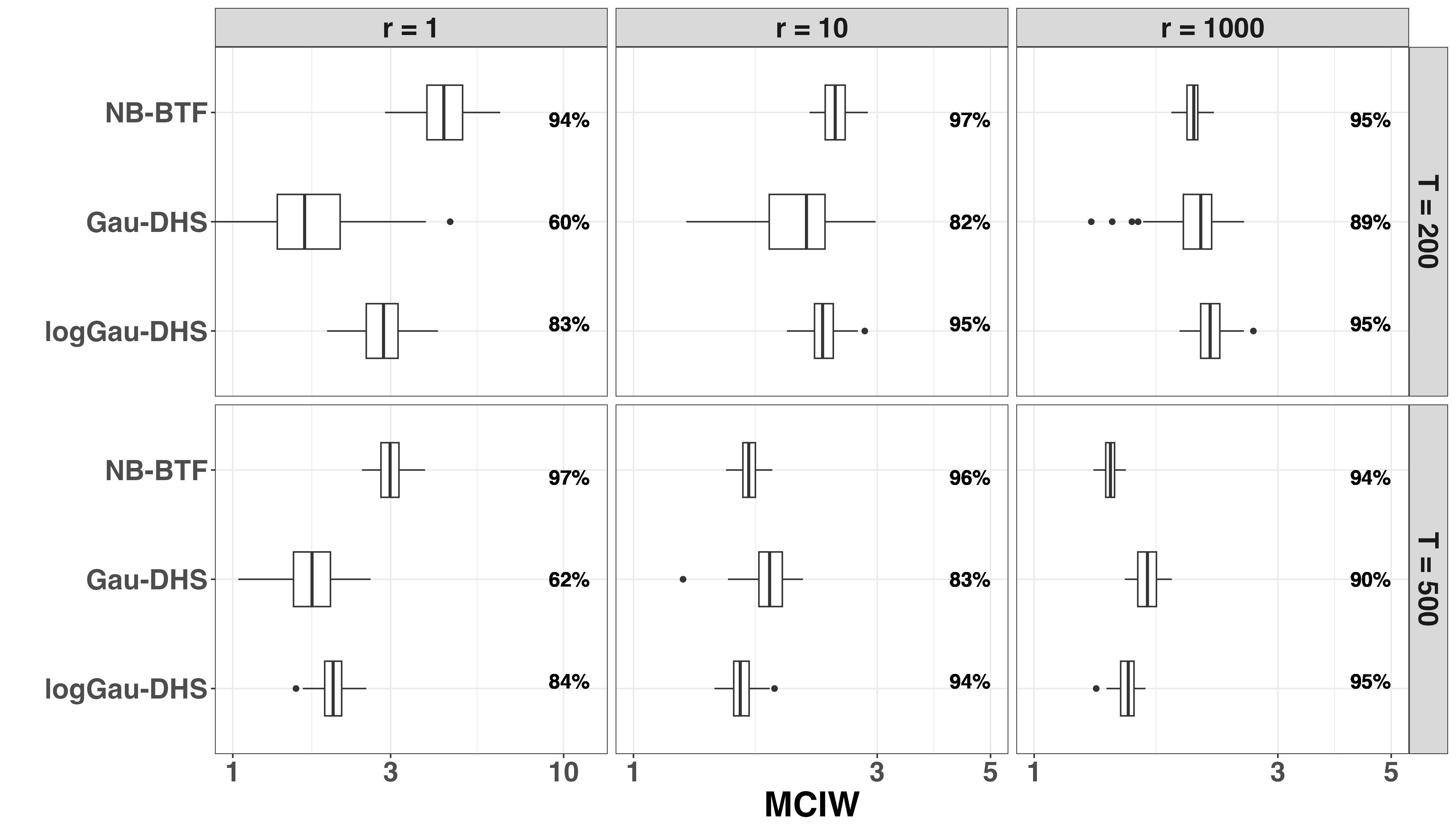}
    \caption{Boxplots of mean credible interval width (MCIW) and average empirical coverage for 95\% equal tail credible intervals for estimation of the true trend under 100 simulations with varying time series length, $T$, and overdispersion, $r$. Models with credible intervals included NB-BTF, Gau-DHS, and logGau-DHS. The x-axis is transformed by log10.}
    \label{fig:sim_mciw}
\end{figure}

The NB-BTF on average has the best accuracy as measured by RMSE across the evaluated simulation settings (Figure \ref{fig:sim_rmse}). The performance is more drastic for the lower density of data setting (T = 200) and more overdispersion (r = 1). The better performance for more overdispersion is likely a result of specifying the true distribution model for the simulated data. However, it is encouraging that the local adaptivity provided by the dynamic horseshoe prior in combination with the correct integer-valued data model is robust to thinning of observations from T = 500 to T = 200.

\begin{figure}[ht!]
    \centering
    \includegraphics[width=\textwidth]{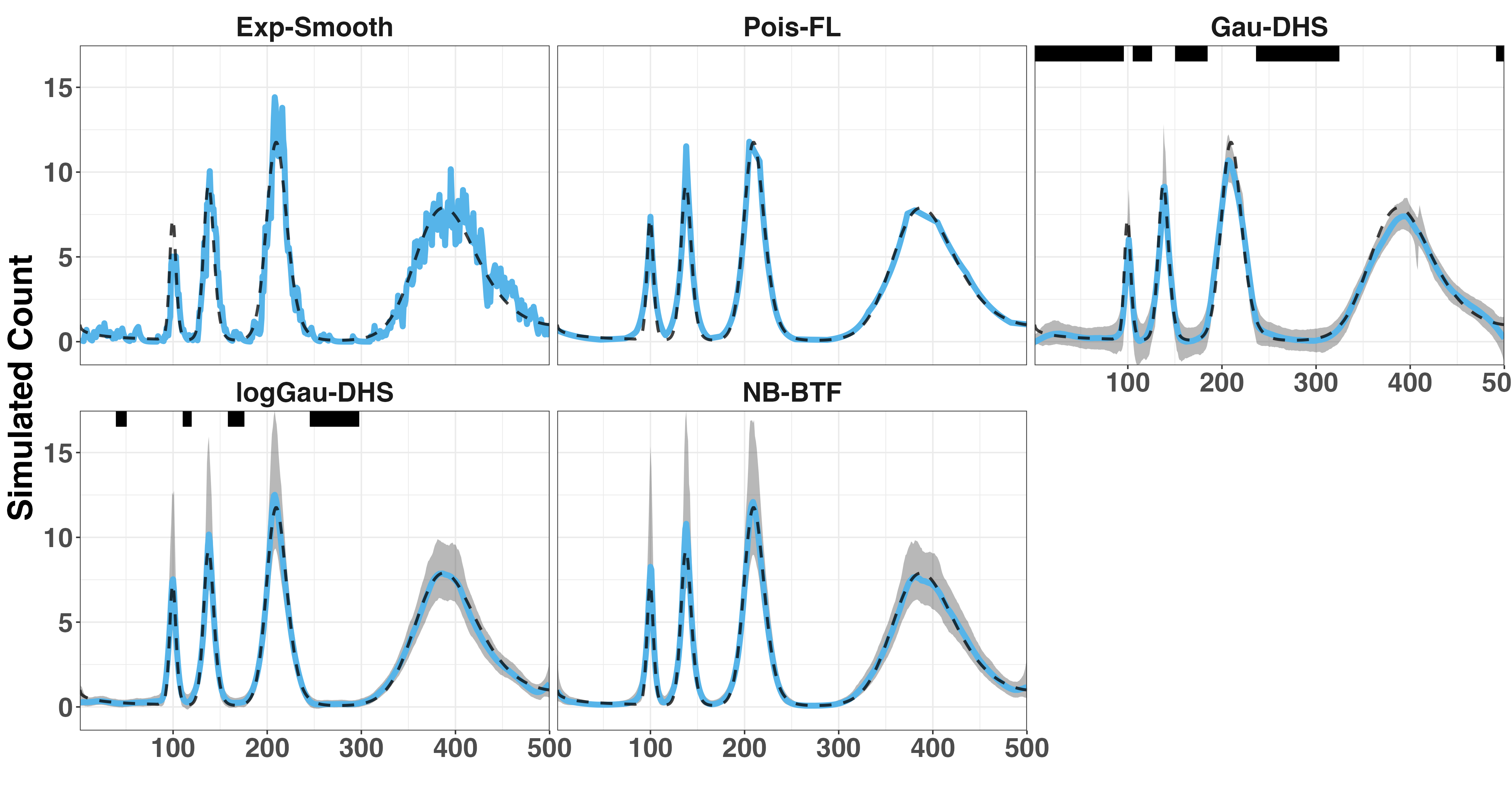}
    \caption{The trend estimates with the smallest RMSE of the 100 simulations with $T=500$ and $r=10$ for each model. The solid blue line is the point estimate of the trend and the dashed black line is the true simulated trend or conditional mean. For Bayesian models, the gray band is the pointwise equal tail 95\% credible interval for the trend. The black bands at the tops of the panels indicate time points in which the credible interval includes a negative lower bound.}
    \label{fig:sim_ts}
\end{figure}

When comparing MCIW, the results were more mixed (Figure \ref{fig:sim_mciw}). While the NB-BTF does not consistently have the narrowest intervals, it does have consistent behavior in the empirical coverage making it the most reliable method for inference. The Gaussian model greatly underperforms in the empirical coverage for majority of the simulation settings. From examining the performance of the best fits in Figure \ref{fig:sim_ts}, it can be seen that that Gaussian estimates are smoother than the truth most likely from the separation of the mean and variance; in other words, the added flexibility of the observation error may contribute to mean structure being incorrectly attributed to error variance. As the data becomes more Poisson-like (r = 1000), the performance of the logGau-DHS improved (Figure \ref{fig:sim_mciw}).

We visually compared the fits of the models in Figure \ref{fig:sim_ts}. It is clear the Exp-Smooth chased the noise in the data. The Pois-FL estimated a trend with multiple sharp features due to the inclusion of exact zeroes in the increments. The Bayesian methods produce smoother trends overall. As mentioned, the Gau-DHS estimates appear to be overly smooth. Additionally, a large proportion of negative lower credible interval bounds were estimated for time periods with a conditional mean close to zero. While it may seem the logGau-DHS and NB-BTF estimates are indistinguishable, the transformation does not guarantee valid inference as several lower credible interval bounds were estimated to be negative. In conclusion, the simulation experiments showed the general superior performance of using the NB-BTF. 

% \section{Applications}

\section{Power Outages}\label{sec_power}

\begin{figure}[ht!]
     \centering
     \begin{subfigure}[b]{0.99\textwidth}
        \centering
        \includegraphics[width = \textwidth]{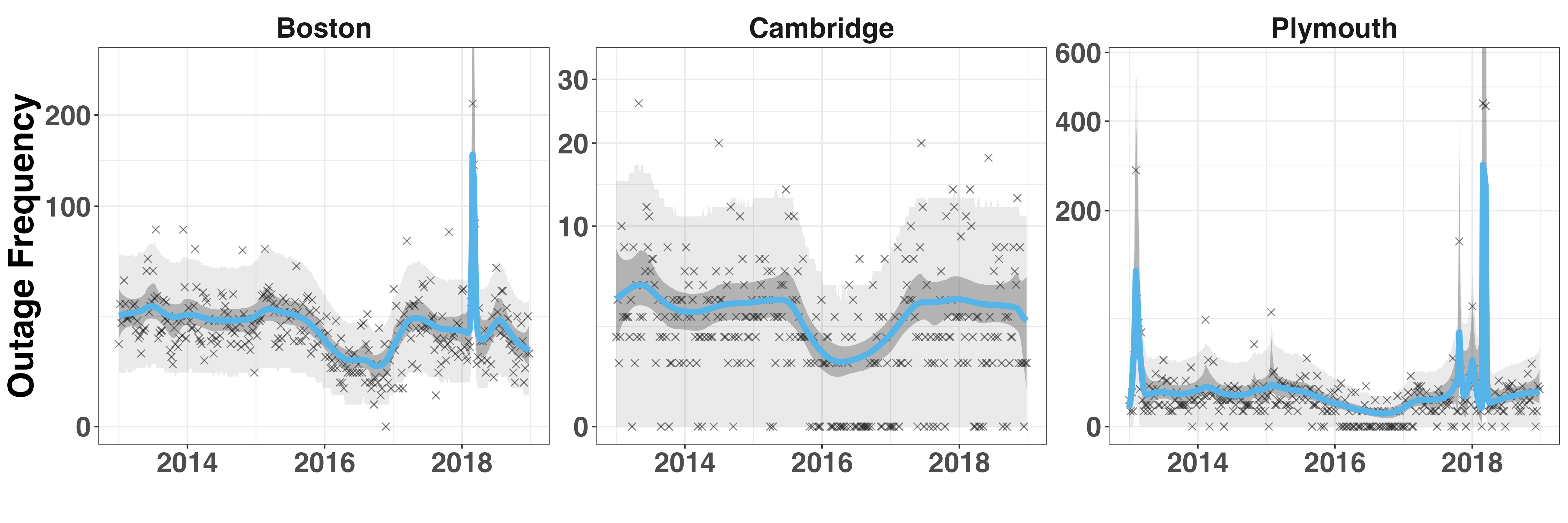}
         \caption{}             
         \label{fig:out-fit}
     \end{subfigure}
      \begin{subfigure}[b]{0.99\textwidth}
        \centering
        \includegraphics[width = \textwidth]{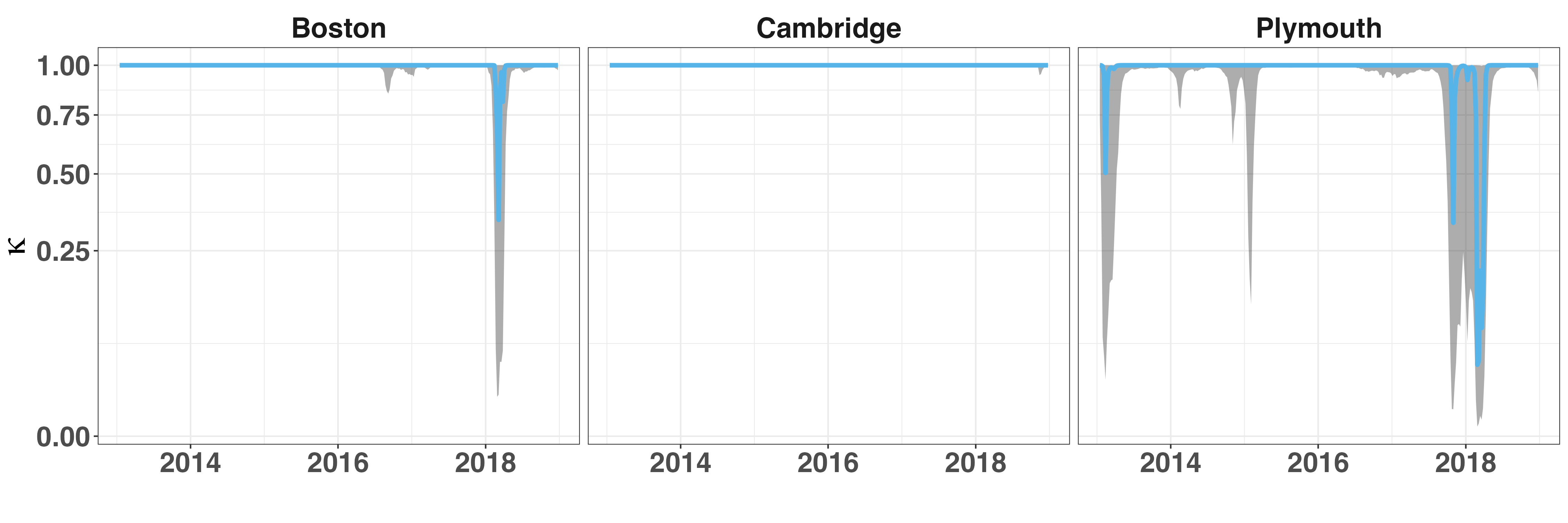}
         \caption{}             
         \label{fig:out-shrink}
     \end{subfigure}
     \caption{(a) The estimated mean trend, blue line, for weekly counts of power outages shown on a square-root scale in Boston, Cambridge, and Plymouth, Massachusetts. (b) The estimated shrinkage, blue line, for weekly counts of power outages in Boston, Cambridge, and Plymouth, Massachusetts. For all plots, the dark grey bands indicate the 95\% equal tail credible interval for the mean trend while the lighter grey bands indicate the 95\% equal tail prediction interval. The x's represent the observed values of the counts. All summaries are point-wise.}
     \label{fig:outage-pred}
\end{figure}

The measure of reliability and resiliency of the electric energy grid in the United States is an active area of research \citep{su2020power,checastaldo2021critical}. Interruptions to the power grid are often characterized by high impact, low frequency events. To demonstrate our framework, we analyzed the weekly aggregated frequency of power outages over the period 2013-2018 in the townships of Boston, Cambridge, and Plymouth, Massachusetts (Figure \ref{fig:outage-pred}) from power outage records downloaded from \citet{CommonwealthofMassachusetts2020}. The frequency refers to the number of outages reported with a start time in the seven day period (week) since the first day of observations, January 1, 2013. The three townships illustrate the model in settings with different nominal levels of counts, number of abrupt jumps, and presence of zeroes. A value of zero corresponds to a week in which no outages started in the corresponding seven day period. In the Boston time series, a zero occurred once during late November 2016 whereas zeroes make up approximately 15\% of the observations in Cambridge and Plymouth. The Boston time series is characterized by a nominal level of outages per week around 25 while Cambridge and Plymouth have nominal levels around 5. Boston and Plymouth time series are characterized by infrequent spikes or bursts in the counts, with the most significant spike around 2018, but no abrupt changes are apparent in the Cambridge power outages. All models were fit to second differences as seasonality is expected to lead to drift in the time periods not affected by shift. 

The estimated trends for the outages capture the features of the data well. The proportion of time estimated to be unshrunk for the three townships is $8/311 \approx 0.026$, $0/311$, and $44/311 \approx 0.141$ for Boston, Cambridge, and Plymouth respectively (Figure \ref{fig:out-shrink}). Therefore, the trend estimation suggests that for a majority of the time the mean number of power outages is evolving persistently with infrequent jumps in the trend for Boston and Plymouth only (Figure \ref{fig:outage-pred}). 

The estimated trend along with the shrinkage profile characterizes a historical analysis of the grid system reliability. The periods of unshrunk estimates indicate periods of higher exposure to risk to the energy grid as they indicate deviation from the typical trend evolution. For instance, the period of time corresponding to a spike in outages in March 2018 was the result of multiple winter storms called nor'easters \citep{liberto2018noreaster}. One of the estimated quantities along with the uncertainty may serve as a \textit{critical risk indicator} for the grid reliability \citep{checastaldo2021critical} and can subsequently be used in a model such as the Bayesian spillover graph of \citet{deng2022bayesian} for quantifying  temporal associations between climate and energy grid risk more formally and strengthening the information used for planning and managing the grid. 

\section{Discussion and Future Work}\label{sec_disc}

The dynamic horseshoe prior for the negative binomial data likelihood provides a flexible integer-valued model for non-stationarity in count time series characterized by low counts. It produces smooth estimates for the temporal trend and integer-valued inference for desired quantities. Furthermore, the specification of negative binomial had consistent performance across a range of settings compared to the Gaussian and log-Gaussian counterparts.  

The model can be used in other application settings such as epidemiology and social science. In epidemiology, inference for the average number of cases at a short temporal and small spatial resolution such as county level COVID-19 cases is likely to be of interest for practitioners, but characterized by smaller counts and zeros. In the social sciences, the study of the frequency of violent events informs social scientists on policy and security decisions. 

One immediate extension for applicability in the mentioned domains is to a spatiotemporal shrinkage process. While we have discussed the temporal shrinkage process, a spatial shrinkage process was introduced by \citet{jhuang2019spatial}. An initial spatiotemporal shrinkage process will likely be defined as separable in space and time as it is a natural starting point.

Additionally, as the prior is part of a hierarchical Bayesian framework, extensions to the continuous DHS such as outlier flagging and changepoint detection \citep{wu2020adaptive} can be adopted to the integer-valued settings. In order to incorporate the extensions, one possibly needs to derive the estimators under the transformation induced by the link function. Ultimately, the goal of future work is to derive a generalized framework for data models in the exponential family to extend to other characteristic models of counts and models for proportion or binary data.

\bigskip
\begin{center}
{\large\bf SUPPLEMENTARY MATERIAL}
\end{center}

\begin{description}

\item[Supplement:] A description of MCMC sampling steps for DHS parameters; computation time, empirical coverage at additional nominal credible interval levels, and a long table of numerical summaries of simulations. (.pdf)

\end{description}

\bibliographystyle{jasa3}

\end{document}

\section{MCMC Sampling Details}\label{supp:mcmc}

The sampling details for the parameters associated with the evolution of the log-volatilities, $\bh = (h_1,...,h_{T-D})'$, are derived in \citet{kowal2019dynamic}, but we provide them again here for completeness. 

\subsection{Log-volatilities}

Using the result of Theorem 4 of \citet{kowal2019dynamic}, the innovation distribution of the evolution equation, $\eta_t \sim \textrm{Z}(1/2, 1/2, 0, 1)$, can be expressed equivalently as the scale mixture:

\begin{equation}\label{eq:innov}
\begin{split}
    \eta_t \vert \xi^{\eta} \sim \textrm{N}(0, (\xi_t^{\eta})^{-1}),\\
    \xi^{\eta}_t \sim \textrm{PG}(1, \eta_t).
\end{split}
\end{equation}
From this result, it follows that $h_t = \mu + \phi(h_{t-1}  - \mu) + \eta_t, \ \eta_t \sim \textrm{N}(0, (\xi_t^{\eta})^{-1})$. 

The conditional posterior distribution of $h_t$ relies on an approximation to the log$\chi^2_1$ given by \citet{omori2007stochastic} and the aforementioned ``all without a loop" (AWOL) algorithm in \citet{kastner2014ancillarity}. From the definition in text, $\omega_t \sim \textrm{N}(0, e^{h_t})$. Define $\tilde{y}_t = \textrm{log}(\omega_t^2 + c)$ for some small constant $c$. Then approximately, $\tilde{y}_t = m_{s_t} + h_t + \epsilon_t, \ \epsilon_t \sim \textrm{N}(0, v_{s_t})$, where $s_t$ is an indicator for one of ten components in a ten component Gaussian mixture model from \citet{omori2007stochastic} with mixture component mean, $ m_{s_t}$, and variance $v_{s_t}$. We do the update of $\boldsymbol{h}$ with a non-centered parameterization by defining $\tilde{h}_t = h_t - \mu$. Then, we have the following state equations:

\begin{equation}\label{eq:vol_state}
    \begin{split}
        \tilde{y}_t = m_{s_t} + \tilde{h}_t + \mu + \epsilon_t, \ \epsilon_t \sim \textrm{N}(0, v_{s_t}),\\
        \tilde{h}_t = \phi\tilde{h}_{t-1} + \eta_t, \ \eta_t \sim \textrm{N}(0, (\xi_t^{\eta})^{-1}),\\
        \tilde{h}_1 \sim \textrm{N}(0, (\xi_1^{\eta})^{-1}).
    \end{split}
\end{equation}

We can jointly update the entire state vector $\boldsymbol{\tilde{h}} = (\tilde{h}_1,...,\tilde{h}_T)'$ by deriving the multivariate conditional Gaussian distribution $\boldsymbol{\tilde{h}} \sim \textrm{N}(\boldsymbol{Q}_{\tilde{h}}^{-1}\boldsymbol{\ell}_{\tilde{h}},\boldsymbol{Q}_{\tilde{h}}^{-1})$ where:

\begin{equation}\label{vol_awol}
    \begin{split}
        \boldsymbol{Q}_{tt,\tilde{h}} = \frac1{v_{s_t}} + \xi^{\eta}_t + \phi^2\xi^{\eta}_{t+1} ,\\
        \boldsymbol{Q}_{t,t-1,\tilde{h}} = -\phi\xi^{\eta}_{t},\\
        \boldsymbol{Q}_{TT,\tilde{h}} = \frac1{v_{s_T}} + \xi^{\eta}_T, \\
        \ell_{t,\tilde{h}} =  \frac1{v_{s_t}}(\tilde{y}_t - m_{s_t} - \mu).
    \end{split}
\end{equation}

\subsection{DHS Parameters}

The global variance, $\tau$, is sampled through the transformation to $\mu$ with aPG parameter expansion. Specifically, the prior $\tau \sim C^+(0,\sigma_\tau)$ implies the transformed prior $\mu \vert \xi_\mu \sim \textrm{N}(\textrm{log}(\sigma^2_\tau), (\xi^\mu)^{-1})$ with $\xi^\mu \sim \textrm{PG}(1,0)$. The conditional update equation is derived by initializing $h_1 \sim \textrm{N}(\mu, (\xi_1^\eta)^{-1})$ with $\xi_1^\eta \sim \textrm{PG}(1,0)$ then $\mu \sim \textrm{N}(Q_\mu^{-1}l_\mu, Q_\mu^{-1})$ with $Q_\mu = \xi^\mu + \xi_1^\eta + (1 - \phi)^2 \sum_{t = 2}^{T} \xi_t^\eta$ and $\ell_\mu = \xi_\mu \textrm{log}(\sigma^2_\tau) + \xi_1^\eta h_1 + \sum_{t = 2}^T \xi_t^\eta (1 - \phi)(h_t - \phi h_{t-1})$.

The autocorrelation, $\phi$, for the dynamic shrinkage process is sampled with slice sampling \citep{neal2003slice} given the above reparameterization of $h$ to $\tilde{h}$ and the PG parameter expansion. Slice sampling requires the full conditional distribution:

\begin{equation}\label{eq:phi}
    \begin{split}
        [\phi \vert \cdot] \propto [\textbf{h} \vert \cdot][\phi],\\
\propto \left(\frac{\phi+1}{2}\right)^{9}\left(\frac{1-\phi}{2}\right) [\tilde{h}_{1} \vert \xi^\eta_{1}] \prod_{t>1} [\tilde{h}_{t} \vert \tilde{h}_{t-1}, \phi, \xi^\eta_{t}],
    \end{split}
\end{equation}
where $\tilde{h}_{1} \vert \xi^\eta_{1} \sim \textrm{N}(0,(\xi^\eta_{1})^{-1})$ and $\tilde{h}_{t} \vert \tilde{h}_{t-1}, \phi, \xi^\eta_{t} \sim \textrm{N}(\phi\tilde{h}_{t-1},  (\xi^\eta_{t})^{-1})$.

\newpage

\section{Simulation Results}\label{supp:sim}

\subsection{Computation Time}

\begin{figure}[ht!]
    \centering
    \includegraphics[width = \textwidth]{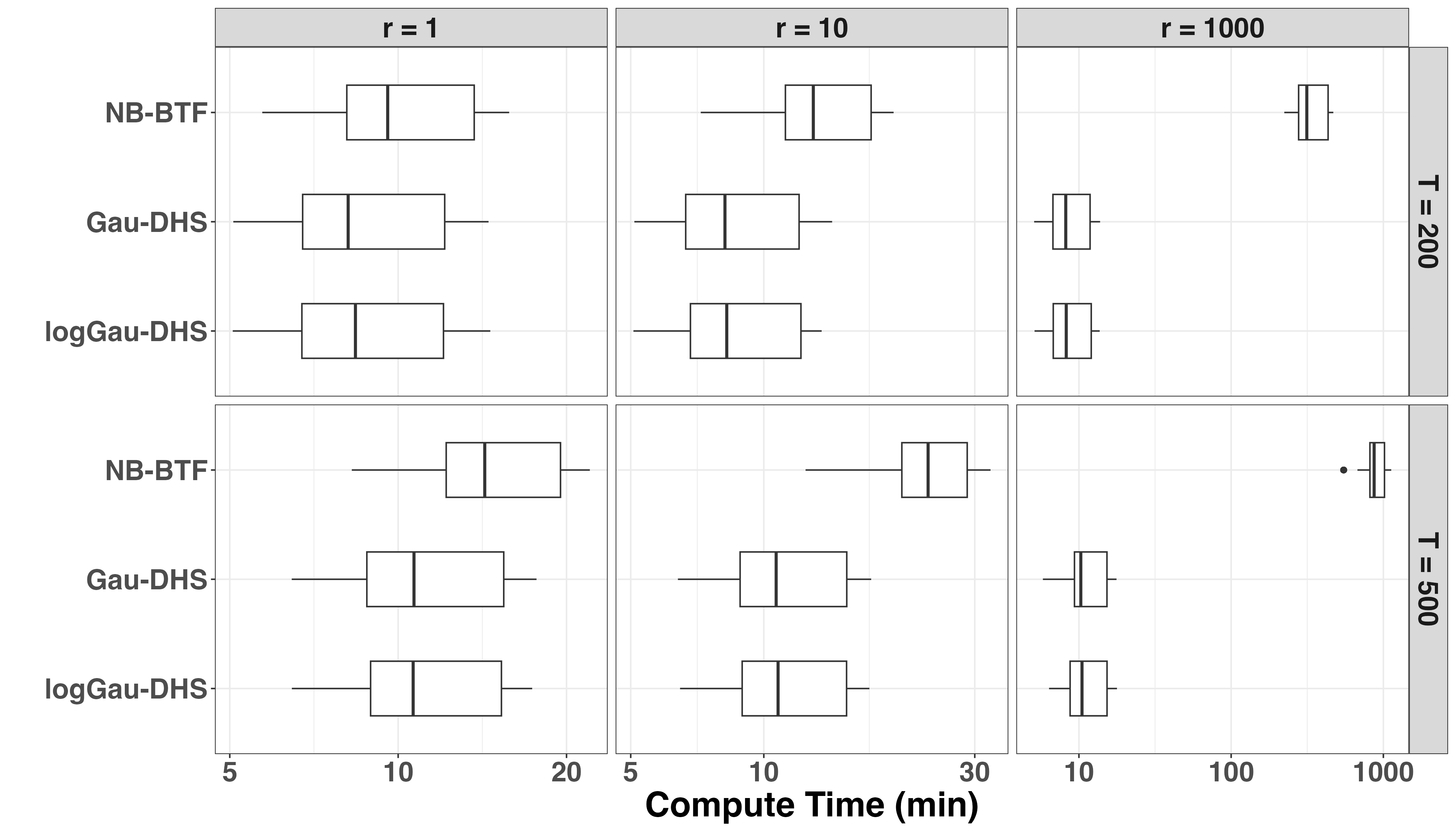}
    \caption{Boxplots of computation time in minutes for the three Bayesian methods considered: NB-DHS, Gau-DHS, and logGau-DHS. All methods have shorter runtimes for smaller datasets. The runtime for the Negative Binomial Bayesian trend filter increases as overdispersion increases. The x-axis is transformed by log10.}
    \label{fig:sup-comp}
\end{figure}

The computation time was recorded for the estimation of the trend for each model compared in the simulation study. The range of computation times for exponential smoothing was 0.002 s to 1.111 s and the distribution was relatively invariant across across the simulation settings. The range of computation times for the Poisson fused lasso was 0.117 s to 0.828 s with faster computation times for fewer time points. The Bayesian methods required significantly more time as shown in Figure \ref{fig:sup-comp}. All the Bayesian methods were run for 105000 iterations with the first 100000 discarded as burnin and thinned to a posterior sample of 1000. The computation time of the two methods based on the Gaussian likelihood (``Gau-DHS-2" and ``logGau-DHS-2") is invariant to the choice of overdispersion, r. However, the computation for the Bayesian negative binomial trend filtering method is adversely affected by an increase in overdispersion. This is due to the use of the P{\'o}lya-Gamma (PG) data augmentation for the likelihood. The sampling runtime of a PG random variable is linear in its first parameter \citep{polson2013bayesian}; for $\xi_t^\theta$ is the first paramter is integer sum $Y_t + r$ which increases as r increases. 

\clearpage

\subsection{Empirical coverage}

\begin{figure}[ht!]
    \centering
    \includegraphics[width = \textwidth]{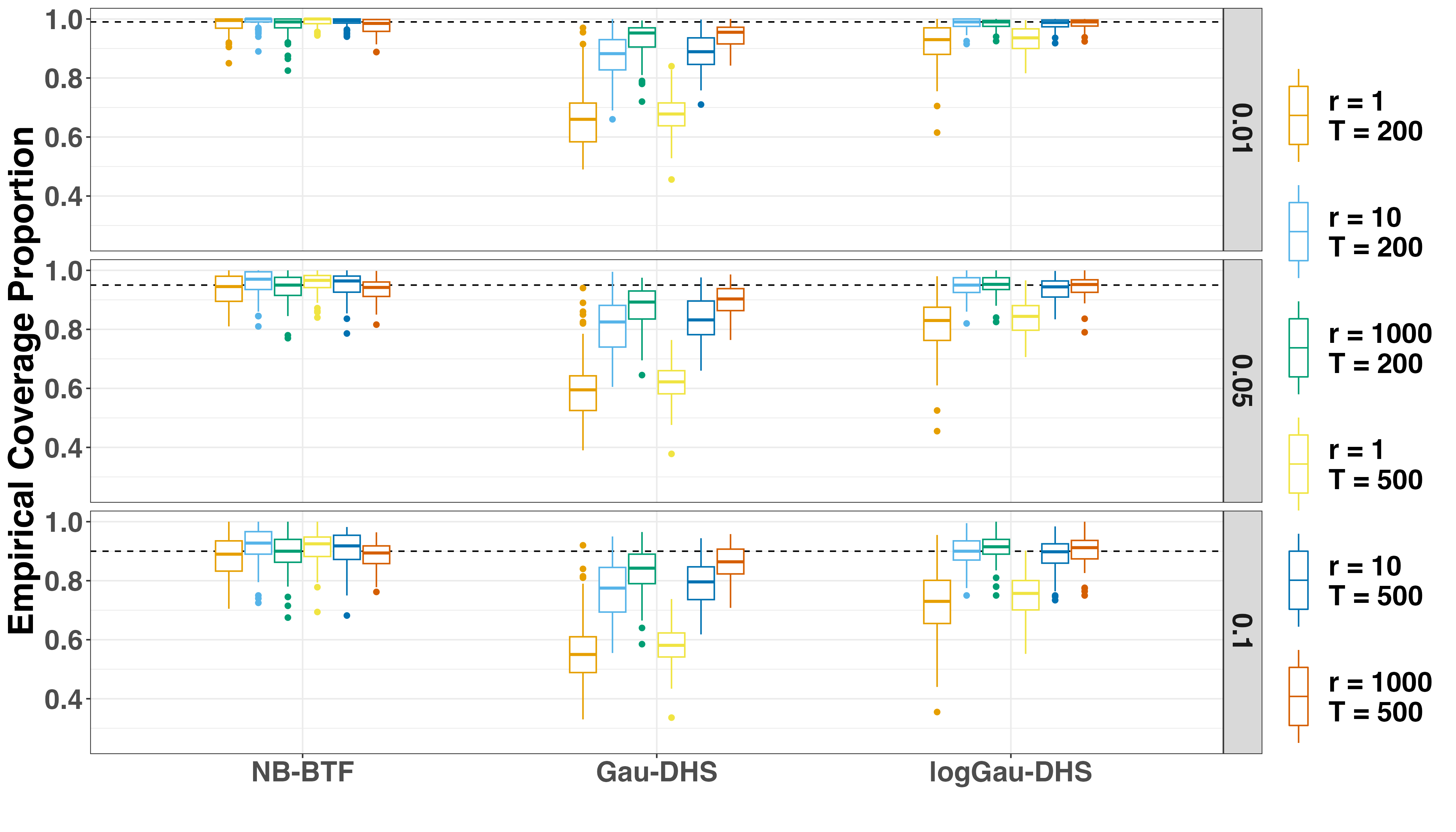}   
    \caption{Boxplots of the empirical coverage proportion for the 0.01, 0.05, and 0.1 equal area credible intervals for the three Bayesian methods considered: NB-DHS, Gau-DHS, and logGau-DHS. The dashed line corresponds to the nominal coverage level. }
    \label{fig:supp-ecov}
\end{figure}

The empirical coverage was calculated across all simulations for the Bayesian methods for equal area credible intervals at the 99\%, 95\%, and 90\% level. The Negative Binomial Bayesian trend filter is the only method to cover the nominal level for all simulation settings (Figure \ref{fig:supp-ecov}). The Gau-DHS and logGau-DHS methods improve coverage as r is increased leading to more Poisson like behavior. However, Gau-DHS is not able to achieve nominal coverage on average across any of the simulation settings considered.

\clearpage

\subsection{Tabulated Results}

The following table reports the mean and standard deviation for root mean squared error (RMSE) and mean credible interval width (MCIW) for 100 simulations of the count data with the trend described in the main text. The variations of the simulation include the time series length (T) and overdispersion (r). The degree of differencing (D) is viewed as a model choice. The six models compared were exponential smoothing (Exp-smooth), Poisson generalized linear model with fused lasso (GLM-L1), our proposed method (NB-DHS), our proposed method with an independent prior ($\phi = 0$; NB-HS), the continuous dynamic shrinkage model of \citet{kowal2019dynamic} on the simulated counts (Gau-DHS), and the continuous dynamic shrinkage on the transformed counts, $\log(Y_t + 1 )$, (logGau-DHS).

\clearpage

\begin{longtable}{rrllll}
\toprule
T & r & Experiment & D & RMSE & MCIW\\
\midrule
\endfirsthead
\multicolumn{6}{@{}l}{\textit{(continued)}}\\
\toprule
T & r & Experiment & D & RMSE & MCIW\\
\midrule
\endhead

\endfoot
\bottomrule
\endlastfoot
 &  & Exp-Smooth &  & 2.31 $\pm$ 0.37 & \\

 &  & Pois-FL & 2 & 2.09 $\pm$ 0.87 & \\

 &  & NB-BTF & 2 & \textbf{1.42} $\pm$ 0.35 & 4.44 $\pm$ 0.73\\

 &  & Gau-DHS & 2 & 4.03 $\pm$ 0.83 & \textbf{1.76} $\pm$ 0.80\\

 & \multirow{-5}{*}{\raggedleft\arraybackslash 1} & logGau-DHS & 2 & 1.65 $\pm$ 0.33 & 2.87 $\pm$ 0.48\\
\cmidrule{2-6}
 &  & Exp-Smooth &  & 1.62 $\pm$ 0.14 & \\

 &  & Pois-FL & 2 & 0.99 $\pm$ 0.30 & \\

 &  & NB-BTF & 2 & \textbf{0.77} $\pm$ 0.16 & 2.49 $\pm$ 0.14\\

 &  & Gau-DHS & 2 & 1.55 $\pm$ 0.40 & \textbf{2.12} $\pm$ 0.43\\

 & \multirow{-5}{*}{\raggedleft\arraybackslash 10} & logGau-DHS & 2 & 0.81 $\pm$ 0.17 & 2.37 $\pm$ 0.17\\
\cmidrule{2-6}
 &  & Exp-Smooth &  & 1.45 $\pm$ 0.12 & \\

 &  & Pois-FL & 2 & 0.79 $\pm$ 0.16 & \\

 &  & NB-BTF & 2 & \textbf{0.66}$\pm$ 0.11 & \textbf{2.05} $\pm$ 0.08\\

 &  & Gau-DHS & 2 & 1.01 $\pm$ 0.28 & 2.09 $\pm$ 0.24\\

\multirow{-15}{*}{\raggedleft\arraybackslash 200} & \multirow{-5}{*}{\raggedleft\arraybackslash 1000} & logGau-DHS & 2 & 0.66 $\pm$ 0.12 & 2.22 $\pm$ 0.15\\
\cmidrule{1-6}
 &  & Exp-Smooth &  & 1.75 $\pm$ 0.22 & \\

 &  & Pois-FL & 2 & 1.35 $\pm$ 0.52 & \\

 &  & NB-BTF & 2 & \textbf{0.99} $\pm$ 0.25 & 2.99 $\pm$ 0.27\\

 &  & Gau-DHS & 2 & 4.06 $\pm$ 0.56 & \textbf{1.73} $\pm$ 0.33\\

 & \multirow{-5}{*}{\raggedleft\arraybackslash 1} & logGau-DHS & 2 & 1.24 $\pm$ 0.23 & 2.02 $\pm$ 0.19\\
\cmidrule{2-6}
 &  & Exp-Smooth &  & 1.14 $\pm$ 0.11 & \\

 &  & Pois-FL & 2 & 0.61 $\pm$ 0.13 & \\

 &  & NB-BTF & 2 & \textbf{0.54} $\pm$ 0.09 & 1.69 $\pm$ 0.07\\

 &  & Gau-DHS & 2 & 1.48 $\pm$ 0.36 & 1.86 $\pm$ 0.16\\

 & \multirow{-5}{*}{\raggedleft\arraybackslash 10} & logGau-DHS & 2 & 0.58 $\pm$ 0.10 & \textbf{1.63} $\pm$ 0.08\\
\cmidrule{2-6}
 &  & Exp-Smooth &  & 1.02 $\pm$ 0.07 & \\

 &  & Pois-FL & 2 & 0.52 $\pm$ 0.11 & \\

 &  & NB-BTF & 2 & \textbf{0.48} $\pm$ 0.08 & \textbf{1.41} $\pm$ 0.04\\

 &  & Gau-DHS & 2 & 0.83 $\pm$ 0.30 & 1.66 $\pm$ 0.09\\

\multirow{-15}{*}{\raggedleft\arraybackslash 500} & \multirow{-5}{*}{\raggedleft\arraybackslash 1000} & logGau-DHS & 2 & 0.48 $\pm$ 0.09 & 1.52 $\pm$ 0.06\\*
\end{longtable}
\clearpage

\bibliographystyle{apalike}